\documentstyle[aps,epsfig]{revtex}                         

\draft
\begin{document}
\twocolumn[\hsize\textwidth\columnwidth\hsize
     \csname @twocolumnfalse\endcsname



\title{Surface relaxations, current enhancements, and
       absolute distances in high resolution STM.
       }

\author{W. A. Hofer, A. J. Fisher}
\address{
         Department of Physics and Astronomy,
         University College, Gower Street, London WC1E 6BT, UK}
\author{R. A. Wolkow}
\address{
         Steacie Institute of Molecular Sciences,
         100 Sussex Drive, Ottawa, Canada K1A 0R6}
\author{P. Gr\"utter}
\address{
         Physics Department, McGill University,
         3600 rue University, Montreal, Canada H3A 2T8}
\maketitle

\begin{abstract}
We have performed the most realistic simulation to date of the
operation of a scanning tunneling microscope.  Probe-sample distances
from beyond tunneling to actual surface contact are covered.  We
simultaneously calculate forces, atomic displacements, and tunneling
currents, allowing quantitative comparison with experimental values.
A distance regime below which the probe becomes unstable is
identified. It is shown that the real distance differs substantially
from previous estimates because of large atomic displacements on the
surface and at the probe-tip.
\end{abstract}


\vskip2pc]

It is hard to avoid the scanning tunneling microscope (STM) in the
imaging, fabrication, or characterisation of nanoscale structures
today
\cite{gimzewski87,clarke96,yanson98,scheer98,olesen96,hallmark87}.
It is certainly the most ubiquitous tool in surface science. But
while the formation of atomic wires from metal leads and their
final breaking is well researched \cite{yanson98,scheer98}, the
physics of the reverse process - an STM tip approaching the
surface at very close range - is far less well documented. This is
so even though the phenomena that occur - the forces, atomic
relaxations, the sudden changes in image at a critical point - are
the everyday experience of every STM experimenter. Even more
importantly, several routes are currently being explored to
fabricate and operate electronic devices on the nanoscale
\cite{eigler90,bartels99,lopinski00}. In chemical methods of
nanofabrication \cite{lopinski00} a high precision of atomic
assembly is attainable, even though the exact position of every
atom is not controlled. By contrast, to initiate such a process,
or to craft structures by moving each element into place, the
exact position of the STM tip and the target atoms must be known.
Furthermore, nanoscale research requires a high degree of
knowledge about the electronic, chemical, and transport properties
of potential constituents - information often obtained by STM.
This information may be misleading if the tip-sample distance is
not correctly estimated. Therefore, as soon as the work moves from
qualitative to quantitative, distance estimates are of critical
importance. The question of distances has previously been analysed
to some extent for STM \cite{clarke96,yanson98,scheer98,olesen96},
and for the related scanning force microscope (SFM)
\cite{jarvis96,perez98,lantz00}. What is missing, so far, is a
picture of the interplay between forces, relaxations, and
currents.

In this paper we analyze the interactions between tip and sample and
their effect on distance estimates and STM images from first
principles. We demonstrate that our simulation technique allows us to
keep track of all key quantities in this situation. We choose the
Au(111) surface since it is widely used in nanotechnology and was
studied in recent extensive experiments \cite{cross98,onishi98}.

\begin{figure}
\begin{center}
\epsfxsize=1.0\hsize \epsfbox{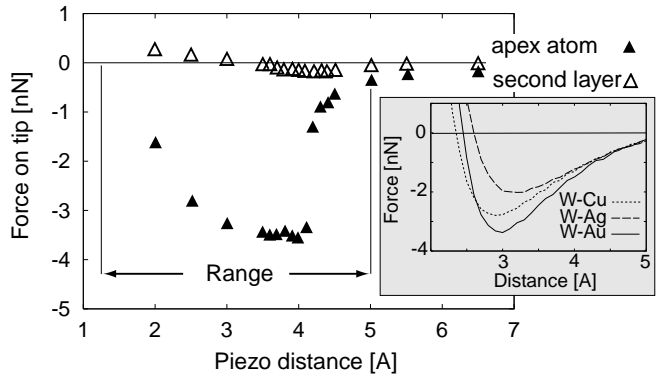}
\end{center}
\caption{
         Forces on the apex atom (full triangles) and the subsurface
         atoms (empty triangles) of the tip above the hollow position
         of the Au(111) surface. The forces on the apex
         atom are about one order of magnitude higher. The range of
         forces is about 4 \AA. Inset: forces on Cu, Ag, and Au atoms
         in a dimer. The forces between the tungsten atom and the
         noble metal atom are roughly equal in a distance range above
         4 \AA.
         }
\label{fig001}
\end{figure}

The computational method is based on first principles density
functional theory. As is standard practice in calculations of this
type only a small part of the periodic surface is used to determine
the physical properties of the whole system. Moreover we employ
periodic boundary conditions both parallel and perpendicular to the
surface. The sample surface was mimicked by a five layer Au(111) film
with nine atoms per layer. The high number of atoms per layer was
necessary for a complete decoupling of the interactions between
adjacent tip pyramids. The central layer of the Au(111) film was our
plane of reference. The tip apex was mimicked by a tungsten
tetrahedron mounted on the reverse of the Au(111) surface slab; the
tip-sample separation was altered by varying the size of our unit
cell normal to the surface. Tungsten is the most commonly used tip
material in experiments; the low reactivity and the similar size of
tungsten and gold atoms makes the support of the tip apex on gold
realistic. We also checked its plausibility by a careful analysis of
the forces during our simulations. In this case the force gradients
are mainly due to the position of an atom, but not its chemical
nature. In Fig. \ref{fig001} we display the forces on the apex atom
and the subsurface atoms of the tip, it can be seen that the forces
on the apex atom are about one order of magnitude higher.

Even for the pseudopotential VASP (Vienna ab initio Simulation
Program \cite{kresse93,kresse96}) code the simulations were rather
demanding and at the limit of realistic time scales. This fact
ruled out a check of the results including additional gold or
tungsten layers. For the same reason we did not include long-range
forces. But as established by Perez et al. in their analysis of
atomic force microscopy \cite{perez98}, they only cause
significant relaxations in the very low distance regime (below 5
\AA), and in this range short-range forces are substantially
higher, as we find in our simulations. The same point was made in
an earlier publication by D\"urig et al. \cite{durig90}. We also
neglect bias-induced surface relaxations in our treatment. The
technical details of the calculation are the following: The cutoff
of the ultrasoft pseudopotentials \cite{vanderbilt90} was 230 eV,
exchange and correlation effects where computed with generalized
gradiant correction and using the formulation of Perdew et al.
\cite{perdew92}. The Brillouin zone was sampled with (4 $\times$ 4
$\times$ 1) k-points of a Monkhorst-Pack grid \cite{monkhorst76},
forces were converged to values of less than 0.01 eV/\AA.

The reading of a piezoscale in the experiments is given by the
position of the plane of reference in the supercell setup of our
system. From an initial tip-sample separation of 10 \AA \, the
distance was gradually reduced, initially in steps of 0.5 \AA. At
every step the system was fully relaxed. The onset of significant
mechanical interactions between tip and sample occurs at about 5.0
\AA. To determine the precise effect of these interactions in the low
distance range we reduced the distance in steps of 0.1 \AA \, from
5.0 to 4.0 \AA \, for one tip position (STM tip on top of a surface
atom), and from 4.5 to 3.5 \AA \, for the second tip position (STM
tip between the surface atoms). We computed the surface wavefunctions
for the fully relaxed sample and tip systems at every distance and
for two tip positions. In this part of the calculation the initial
system was separated into two subsystem, the sample and the tip
surface, and the surface wavefunctions computed for each subsystem
separately. These wavefunctions were used as the input for our
simulation of the tunnel currents \cite{hofer00a}. The procedure
leads to separate constant current contours of the surface, depending
on the position of the tip. Interpolating between the two simulations
we obtain the change of corrugation due to system relaxations from
first principles.

It has been shown by a number of authors
\cite{clarke96,olesen96,perez98} that substantial relaxations are
confined to the low distance regime. Perez et al. obtained their
result with an ab initio method, but they calculated the changes
on a semiconductor surface and did not relate them to measured
currents. Olesen et al., and Clarke et al. analysed the events on
metal surfaces, but employed empirical potentials; for this reason
they too could not simultaneously evaluate the current. Their
results show, not surprisingly, that the onset of forces and
relaxations begins when the interatomic distances are close to the
typical bondlength of the atoms. However, from experiments we know
that tip-sample forces on Au(111) are appreciable for distances of
at least 5 \AA \, \cite{cross98,guggisberg00}. Di Ventra and
Pantelides employed density functional theory to determine the
relaxation of surface atoms for Al(110) surfaces, mimicking the
tip by a single Al atom \cite{diventra99}. Within the range of
distances from the point contact to the tunneling regime (1.3 to
3.3 \AA) they predict no corrugation due to atomic relaxations.
The present calculation is in a sense complementary to their work.
It is shown here that experimental conditions in STM scans do not
allow such a close approach, because the system becomes unstable
already at larger distances (4.6 \AA). Our calculations establish
an early onset of relaxations, which depend strongly on the
position of the tip. For the on-top position of the STM tip we
observe an onset of forces and relaxations at about 4.7 \AA, about
1.0 \AA \, higher than suggested by earlier simulations
\cite{clarke96,olesen96}. The surface atom reaches its furthest
elongation from the surface at a tip-sample separation of 4.5 \AA;
it is then 1.3 \AA \, above its position on the isolated surface.
This point marks the beginning of atomic transfer from the surface
to the tip, the origin of the hysteresis observed in single
approach/retraction cycles \cite{yanson98}. Neighboring atoms on
the surface are equally relaxed, their maximum displacement is 0.1
\AA. The second nearest neighbors remain close to their original
position (vertical displacement less than 0.04 \AA). Even though
the boundary conditions at our two dimensional repeat unit
constrain the relaxations of individual atoms, the very small
relaxation of second nearest neighbors indicates a rapid decay of
the forces in lateral direction. Therefore we do not expect
substantial changes in a calculation performed with a larger unit
cell. If the tip is in the threefold hollow, the onset of forces
and relaxations is substantially retarded and occurs at 4.3 \AA.
The maximum outward relaxation of the three gold atoms is somewhat
lower and about 1.0 \AA \, (reached when the tip-sample separation
is 4.1 \AA).

\begin{figure}
\begin{center}
\epsfxsize=1.0\hsize \epsfbox{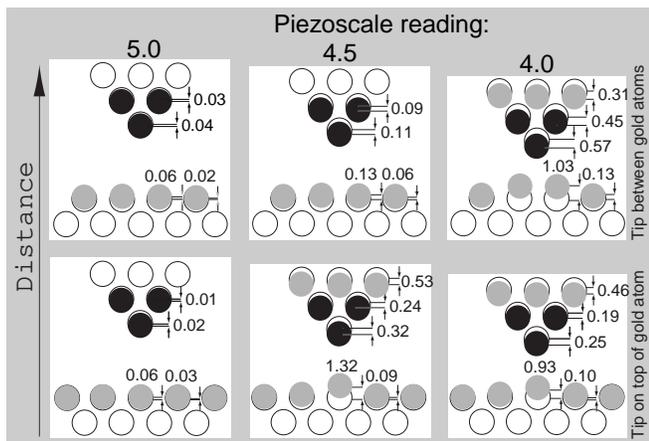}
\end{center}
\caption{
         Relaxed positions of sample and tip
         atoms for three piezoscale readings of the STM:
         5.0 \AA (left), 4.5 \AA (center), 4.0 \AA (right);
         and two STM tip positions. At the point of closest
         approach the closest Au surface atoms are about 1 \AA
         higher than on the isolated surface. Note that
         relaxations differ substantially for the two
         different positions.
        }
\label{fig002}
\end{figure}

\begin{figure}
\begin{center}
\epsfxsize=1.0\hsize \epsfbox{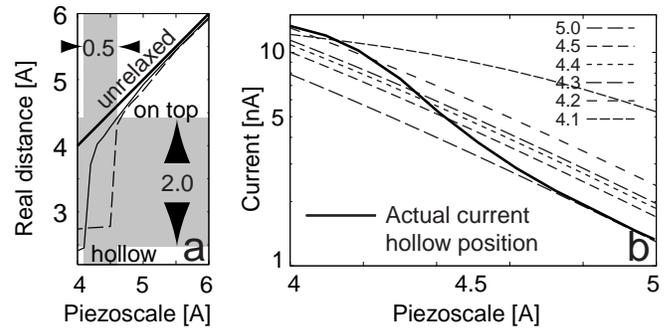}
\end{center}
\caption{
            Relaxations and currents of the coupled system.
            The separation decreases, within a distance of 0.5 \AA \,
            by as much as 2.0 \AA \, (a). Logarithmic current graphs
            computed from the wavefunctions of the relaxed system.
            Individual curves pertain to equilibrium positions at
            a specific piezoscale reading. In the range of
            measurements at about 4.5 - 5.0 \AA \, the current decays
            faster due to relaxation effects, the actual barrier
            height seems increased (b).
         }
\label{fig003}
\end{figure}

The tungsten tetrahedron of the STM tip relaxes about the same amount
in both positions: the apex atom approaches the sample by about 0.5
\AA \, at the point of closest approach. The change is due to an
increase of the interlayer distance of the tungsten atoms by about
0.1 to 0.2 \AA, and a buckling of the layers underneath, which
extends about two layers into the lead. The forces on the tungsten
tip are shown in Fig. \ref{fig001}. The values for tip-sample
relaxations are given in Fig. \ref{fig002}. Owing to the strong
interactions between tip and sample at close range, the core-core
distance between tip and sample is decreased. The net effect of
relaxations is a decrease of the distance between tip and sample. A
deviation of about 2 \AA \, between the distance inferred from
simulations, and the distance inferred from measurements of the
decay, has actually been observed \cite{hofer00b}. In the critical
range of 4.6 to 4.1 \AA \, the actual distance changes by about 2
\AA. The real distances are shown in Fig. \ref{fig003} (a). This
points to an effect frequently observed in experiments with atomic
resolution: it is difficult to obtain atomically-resolved images,
because sizable corrugations in most cases require a distance below 5
\AA, but it is very easy to destroy an STM tip by small increases of
the current. We now understand that the actual range of imaging is
less than 1.0 \AA, or less than an order of magnitude in the current
scale. Stability is destroyed, when the apex atom of the tip and an
atom of the sample form a chemical bond. To check, whether a
different metal surface would lead to different results and thus a
different interval in actual measurements we computed the forces
between a tungsten atom and a Cu, Ag, and Au atom in a dimer
configuration. The result of our calculation is shown in the inset of
Fig. \ref{fig001}. It can be seen that the forces are equal for all
three noble metals in the range above 4 \AA. Short of performing
identical simulations for all three surfaces this seems the best
confirmation that our result is of general validity.

Currents were calculated in a perturbation approach
\cite{hofer00a} and based on the fully relaxed sample and tip
systems. The current was calculated for a sample bias of  -0.1 V.
The curves in Fig. \ref{fig003} (b) show the current for the
hollow position. We obtained the change due to surface and tip
relaxations by calculating the full range of currents in a frozen
atomic configuration. The logarithmic graphs are different for
every distance, the z-labels give the tip-sample separation before
relaxation. The true increase of the current is determined by
taking only a single point of each graph. In practice we
interpolated between the computed points to obtain the smooth
black curve shown in the figure, giving an enhanced increase in
the current. For an experimenter who estimates the potential
barrier between tip and sample from the exponential decay of the
current, the apparent barrier height increases in this range
\cite{durig90,schirmeisen00}. The current calculation is based on
the Baardeen approach \cite{bardeen61}, where sample and tip are
considered decoupled. In a previous paper we studied the effects
of a tip potential on the surface electronic structure
\cite{hofer00b}, and it was established that these effects are
limited to very low distances (below 4 \AA). Here we find that the
onset of relaxations begins about 1 \AA \, before this point. The
main effect, not included in perturbation theory, is therefore due
to relaxations. But this indicates that a more refined treatment
of STM currents e.g. by a scattering approach is no substantial
improvement compared to the Bardeen approximation, if it does not
simultaneously treat relaxations. And this, in turn, is still
computationally untractable because it requires a fully ab initio
treatment for every single position of the tip.

\begin{figure}
\begin{center}
\epsfxsize=1.0\hsize \epsfbox{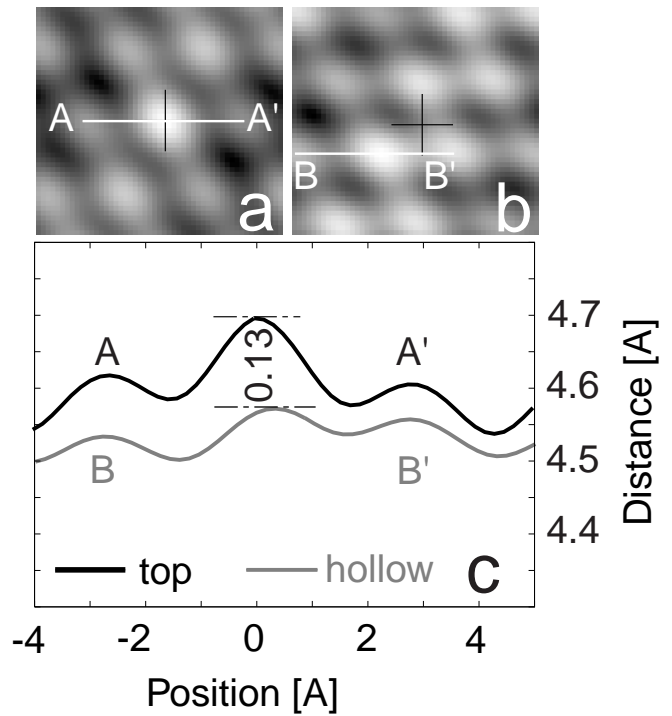}
\end{center}
\caption{
            Current contour plots using the relaxed configurations of sample
            and tip for a distance of 4.7 \AA \, (tip on top of an Au atom,
            frame (a)), and 4.5 \AA \, (tip in the threefold hollow position,
            frame (b)). Simulated linescan across a single atom (c), for
            the setup in the on-top position (black curve, A-A' in frame (a))
            and for the setup in the hollow position (grey curve, B-B' in
            frame (b)). All scans correspond to a current of 5.1 nA.
            An actual scan would have enhanced corrugation due to
            relaxation effects (see dash dotted lines in frame (c)).
         }
\label{fig004}
\end{figure}

All the results presented here are derived from ab initio
treatments of currents and relaxations for two lateral positions
of the tip: the on-top position and the threefold hollow. The
tunnel currents in our simulations were assumed to be 5.1 nA in
both cases, but owing to the site-dependent relaxation of the
systems the two constant-current contours differ. The piezoscale
would give a distance of 4.7 \AA \, on top of a gold atom and 4.5
\AA \, in the fcc hollow site. These distances are at the lowest
limit of stability for the tip-sample system. The difference of
0.2 \AA \, describes the experimental corrugation of the Au(111)
surface, when measured with a tungsten terminated tip
\cite{hallmark87}. The simulated scans with the tips and surfaces
frozen in the two relaxed configurations are shown in Fig.
\ref{fig004} (a), and (b). We stress that these images are
idealizations, because they were computed with the wavefunctions
of frozen atomic positions of tip and sample. In Fig. \ref{fig004}
(c) we show two simulated linescans across a single atom. The two
separate curves describe the linescan for the two setups used
(on-top, and hollow). It can be seen that the relaxation leads in
effect to an enhancement of the corrugation by about 0.13 \AA.
Only with this relaxation do we obtain agreement between
experiments and simulations \cite{hallmark87}.

The most remarkable result of our calculations, and the one that
experimenters may find most important, is the very low range of
stability for atomically resolved STM scans on metal surfaces. This
range is only about 0.5 \AA. Below this point, at about 4.5 \AA \,
from the surface, the distance between tip and sample changes
drastically by 2 \AA \, (within 0.5 \AA \, of measured displacement),
as the two separate systems jump into contact. In this range the
current may even decrease as the tip approaches, if one atom of the
tip or sample system becomes separated.

The work was supported by the British Council and the National
Research Council. Computing facilities at the UCL HiPerSPACE center
were funded by the Higher Education Funding Council for England. AJF
was also supported by an Advanced Fellowship from the Engineering and
Physical Sciences Research Council. PG was supported by the National
Science and Engineering Research Council of Canada and Le Fonds pour
la Formation de Chercheurs et l'Aide a la Recherche de la Province de
Quebec.

%

\end{document}